\documentstyle[12pt,amsfonts]{article}
\parskip=\medskipamount

\def\?{\hglue1\parindent\ignorespaces} % Our convention
\newcommand {\cM}{{\cal M}}

\newcommand {\cR}{{\cal R}}

\def\a{\alpha}

\def\d{\delta}

\def\G{\Gamma}

\def\s{\sigma}

\def\F{\Phi}
\def\J{\Psi}
\def\L{\Lambda}

\def\S{\Sigma}
\def\U{\Upsilon}

\newcommand{\ad}{{\dot{\alpha}}}                       %new
\newcommand{\pa}{\partial}                           %new
\newcommand{\hf}{\frac12}

\newcommand{\be}{\begin{equation}}
\newcommand{\ee}{\end{equation}}
\newcommand{\bea}{\begin{eqnarray}}
\newcommand{\eea}{\end{eqnarray}}

\newcommand{\beq}{\begin{equation}}
\newcommand{\eeq}{\end{equation}}

\font\ro=cmsy10       % font with rope
\def\kcr{{\hbox{\ro \char'170}}}    % right-handed rope
\def\ktl{{\hbox{\ro \char'170}}}  % top end for left-handed rope
\def\ktr{{\hbox{\ro \char'170}}}  % " right
\def\kbl{{\hbox{\ro \char'170}}}  % " bottom left
\def\kbr{{\hbox{\ro \char'170}}}  % " right

\def\endtitle{\end{quotation}\newpage}     % end title page

\def\headpic{           % UM heading
  \indent
  \setlength{\unitlength}{.4mm}
  \thinlines
  \par
  \begin{picture}(29,16)
  \put(165,16){\line(1,0){4}}
  \put(170,16){\line(1,0){4}}
  \put(180,16){\line(1,0){4}}
  \put(175,0){\line(1,0){4}}
  \put(180,0){\line(1,0){4}}
  \put(185,0){\line(1,0){4}}
  \put(169,0){\line(0,1){16}}
  \put(170,0){\line(0,1){16}}
  \put(179,0){\line(0,1){16}}
  \put(180,0){\line(0,1){16}}
  \put(184,0){\line(0,1){16}}
  \put(185,0){\line(0,1){16}}
  \put(169,16){\oval(8,32)[bl]}
  \put(170,16){\oval(8,32)[br]}
  \put(179,0){\oval(8,32)[tl]}
  \put(185,0){\oval(8,32)[tr]}
  \end{picture}
  \par\vskip-6.5mm
  \thicklines}

\def\border{           % border
  \setlength{\unitlength}{1mm}
  \newcount\xco
  \newcount\yco
  \xco=-21
  \yco=12
  \begin{picture}(140,0)
  \put(\xco,\yco){$\ktl$}
  \advance\yco by-1
  {\loop
  \put(\xco,\yco){$\kcr$}
  \advance\yco by-2
  \ifnum\yco>-240
  \repeat
  \put(\xco,\yco){$\kbl$}}
  \xco=158
  \yco=12
  \put(\xco,\yco){$\ktr$}
  \advance\yco by-1
  {\loop
  \put(\xco,\yco){$\kcr$}
  \advance\yco by-2
  \ifnum\yco>-240
  \repeat
  \put(\xco,\yco){$\kbr$}}
  \put(-20,13){\tiny University of Maryland Elementary Particle
Physics University of Maryland Elementary Particle Physics University of
Maryland Elementary Particle Physics}
  \put(-20,-241.5){\tiny University of Maryland Elementary
Particle Physics University of Maryland Elementary Particle Physics
University of Maryland Elementary Particle Physics}
  \end{picture}
  \par\vskip-8mm}

\catcode`@=11    % WATCH OUT !!!!! This must be reset in the end !!!!!
\newbox\t@b@x
\def\rightarrowfill{$\m@th \mathord- \mkern-6mu
     \cleaders\hbox{$\mkern-2mu \mathord- \mkern-2mu$}\hfill
      \mkern-6mu \mathord\rightarrow$}
\def\leftarrowfill{$\m@th \mathord\leftarrow \mkern-6mu
     \cleaders\hbox{$\mkern-2mu \mathord- \mkern-2mu$}\hfill
      \mkern-6mu \mathord-$}
\def\leftrightarrowfill{$\m@th \mathord\leftarrow \mkern-6mu
     \cleaders\hbox{$\mkern-2mu \mathord- \mkern-2mu$}\hfill
      \mkern-6mu \mathord\rightarrow$}
\def\tooo#1{\setbox\t@b@x=\hbox{\footnotesize$#1$}%
             \mathrel{\mathop{\hbox to\wd\t@b@x{\rightarrowfill}}%
              \limits^{#1}}\,}
\def\froto#1{\setbox\t@b@x=\hbox{\footnotesize$#1$}%
             \mathrel{\mathop{\hbox to\wd\t@b@x{\leftrightarrowfill}}%
              \limits^{#1}}\,}
\catcode`@=12                   % You see ?

\begin{document}
\border\headpic {\hbox to\hsize{March 1999 \hfill {UMDEPP 99-091}}}
\par
\setlength{\oddsidemargin}{0.3in}
\setlength{\evensidemargin}{-0.3in}
\begin{center}
\vglue .03in

{\large\bf\boldmath 4D, $N$ = 2 Supersymmetric Off-shell $\s$-Models \\
on the Cotangent Bundles of K\"ahler Manifolds}\footnote{Supported in
part by US NSF Grant PHY-98-02551, the US DOE Grant DE-FG02-94ER-40854,
\\ \?\? and the Deutsche Forschungsgemeinschaft.}${}^,$\footnote{Contribution
to the proceedings of the Buckow-98 meeting of presentation by S.~Kuzenko.}
\\[.3in]
S.~James Gates, Jr.\\[-1mm]
{\it Department of Physics,
University of Maryland at College Park \\[-1mm]
College Park, MD 20742-4111, USA\/}\\[-1mm]
{\tt gates@bouchet.physics.umd.edu}\\[0.2in]
and
\\[0.1in]

Sergei M.~Kuzenko \footnote{On leave from  the
Department  of Physics, Tomsk State University, Russia} \\ [-1mm]
{\it Institut f\"ur Theoretische Physik,  Universit\"at
M\"unchen\\[-1mm]
Theresienstr. 37, D-80333 M\"unchen, Germany\/}\\[-1mm]
{\tt Sergei.Kuzenko@Physik.Uni-Muenchen.DE}\\[0.2in]

{\bf ABSTRACT}\\[.002in]
\end{center}
\begin{quotation}
{We review the construction of 4D $N=2$ globally supersymmetric
off-shell nonlinear sigma models whose target spaces are
the cotangent bundles of K\"ahler manifolds.
}\endtitle

A classical result of SUSY field theory in four spacetime 
dimensions is that the target spaces of rigid supersymmetric 
nonlinear sigma models must be K\"ahler manifolds \cite{zumino} 
when $N=1$ and hyperk\"ahler manifolds  \cite{agf} for $N=2$. The
hyperk\"ahler manifolds form a subspace in the family of K\"ahler
manifolds of complex dimension $2n$ and possess more restrictive 
properties as compared to general K\"ahler ones (three 
independent parallel complex structures, Ricci-flatness, 
$Sp(n)$ holonomy versus $U(2n)$ in general), see \cite{besse} 
for a review.  Remarkably, there exist at least two constructions
which generate hyperk\"ahler manifolds of complex dimension $2n$  
from K\"ahler manifolds of complex dimension $n$.  The first 
example is provided by the famous $c$-map \cite{cfg} in the 
limit of {\it rigid} supersymmetry. One starts with a holomorphic
prepotential $F(Z^i)$, $i=1,2,\dots,n$,  destined to generate 
the so-called rigid special K\"ahler geometry \cite{skg} whose 
potential and metric read
$$
G( Z, \overline{Z}) = F_i (Z)  \overline{Z^i} +
\overline{F}_i (\overline{Z}) Z^i ~,\qquad \qquad
g_{i \bar j} ( Z, \overline{Z}) = F_{ij} (Z) + 
\overline{F}_{ij}(\overline{Z})  \;. 
\eqno(1) $$
Then, it turns out that the following potential
$$
H(Z, \overline{Z}, W,  \overline{W}) = G( Z, \overline{Z}) 
+ \hf \, g^{i \bar j}(Z, \overline{Z}) ( W_i + \overline{W}_i)
( W_j + \overline{W}_j)
\label{cmaphc}
\eqno(2)$$
corresponds to a hyperk\"ahler manifold \cite{cfg}. 
The second example has its origin in the seminal 
paper of Calabi \cite{calabi} where the concept of 
hyperk\"ahler manifolds was introduced. In \cite{calabi}  
Calabi showed that the cotangent bundles of complex 
projective spaces ${\Bbb C}P^n$ are hyperk\"ahler 
manifolds. It was long conjectured and recently 
proved \cite{kaled} that the holomorphic cotangent 
bundles of general K\"ahler manifolds admit 
hyperk\"ahler structures. Therefore, one can 
associate an $N=2$ supersymmetric sigma model with 
any  K\"ahler manifold.  Such manifestly $N=2$ 
supersymmetric sigma models have been described 
in our recent paper \cite{gk}. Below we will present 
a review of our construction.

Let us start by recalling a general $N=1$ rigid supersymmetric 
sigma model \cite{zumino}. The model is described by chiral 
superfields $\F^I$ and their conjugates ${\bar \F}^{\bar I}$ 
whose physical scalar components $A^I$ and ${\bar A}^{\bar 
I}$ parametrize a K\"ahler manifold $\cM$. The action reads
$$
S[\F, \bar \F] ~=~  \int {\rm d}^8 z ~ K(\Phi^{I},
\, {\bar \Phi}{}^{\bar{J}})  
\label{nact4}
\eqno(3)$$
where $K(A, \bar A)$ is the  K\"ahler potential of $\cM$.

An $N=2$ supersymmetric extension of (1), in
which we are here interested, is given in $N=1$ superspace by
$$
S[\U, \breve{\U}] \, = \, \int {\rm d}^8 z \,  \Big[ \, \,
\frac{1}{2\pi {\rm i}} \, \oint \frac{{\rm d}w}{w} \,  \, 
K \big( \U^I (w), \breve{\U}^{\bar{I}} (w)  \big) ~~ \Big]  
~~~.
\label{nact} 
\eqno(4)$$
For $\U (w)$ and $\breve{\U} (w) $ we have
$$ \U^I (w) = \sum_{n=0}^{\infty}  \, \U_n^I (z) w^n = 
\F^I(z) + w \S^I(z) + O(w^2) ~,~~~
\breve{\U}{}^{\bar{I}} (w) = \sum_{n=0}^{\infty}  \, {\bar
\U}_n^{\bar{I}}
(z) (-w)^{-n}
\label{exp}
\eqno(5) $$ 
with $\F$ being chiral, $\S$ being complex linear, and the 
remaining component superfields being unconstrained complex 
superfields.  The superfields $\F$ and $\S$ are constrained 
by 
$$ {\bar D}_\ad \F =0~, \qquad \qquad {\bar D}^2 \S = 0 
\eqno(6) $$
and provide two {\it {distinct}} off-shell realizations of $N=1$ scalar 
multiplets. The role of the auxiliary superfields $\U_2, \U_3, \dots$, 
is to ensure a linearly realized $N=2$ supersymmetry.  The expansions 
in (5) describe ``polar'' multiplets in the nomenclature 
of \cite{grr}. 

The $N=2$ sigma model introduced respects all the geometric features of
its $N=1$ predecessor in (3). The K\"ahler invariance of
(3)
$$ K(\F, \bar \F) \quad \longrightarrow \quad K(\F, \bar \F) ~+~ \Big(
\L(\F)
\,+\,  {\bar \L} (\bar \F) \Big)
\eqno(7) $$
turns into 
$$ K(\U, \breve{\U})  \quad \longrightarrow \quad K(\U, \breve{\U}) ~+~
\Big(\L(\U) \,+\, {\bar \L} (\breve{\U} ) \Big)
\eqno(8) $$
for the model (3). A holomorphic reparametrization $A^I
~\rightarrow~ 
f^I \big( A \big)$ of the K\"ahler manifold has the following
counterparts
$$ \F^I  \quad  \longrightarrow   \quad f^I \big( \F \big) ~, \qquad \qquad 
\U^I (w) \quad  \longrightarrow  \quad f^I \big (\U(w) \big)
\eqno(9) $$
in the $N=1$ and 2 cases, respectively. Therefore, the physical
superfields of the $N=2$ theory
$$ \U^I (w)\Big|_{w=0} ~=~ \F^I ~,\qquad  \quad \frac{ {\rm d} \U^I (w) 
}{ {\rm d} w} \Big|_{w=0} ~=~ \S^I ~,
\label{geo3} 
\eqno(10) $$
should be regarded, respectively, as a coordinate of the K\" ahler
manifold and a tangent vector at point $\F$ of the same manifold. 
That is why the variables $(\F^I, \S^J)$ parametrize the tangent 
bundle $T\cM$ of the K\"ahler manifold $\cM$. 

The presence of auxiliary superfields $\U_2, \U_3, \dots $, in 
(5) makes $N=2$ supersymmetry manifest, but the physical 
content of the theory is hidden. To describe the theory in terms of 
the physical superfields $\F$ and $\S$ only, all the auxiliary 
superfields have to be eliminated  with the aid of the 
corresponding algebraic equations of motion
$$ \oint \frac{{\rm d} w}{w} \,w^n \, \frac{\pa K(\U, \breve{\U} 
) }{\pa \U^I} ~ = ~ \oint \frac{{\rm d} w}{w} \,w^{-n} \, \frac{\pa 
K(\U, \breve{\U} ) } {\pa \breve{\U}^{\bar I} } ~ = ~
0 ~~~, \qquad n \geq 2 ~~~ .               
\label{int}
\eqno(11) $$
Their solution $\U_{\star} (w)$ can be found only perturbatively
for general K\"ahler manifolds. Remarkably, there exist numbers of 
special cses, for instance, the four series of compact K\"ahler 
symmetric spaces  (see, e.g. \cite{besse,perelomov})
%\bea 
$$\frac{ SU(m+n) }{ SU(m) \times SU(n) \times U(1) } ~~,
{~~~~~~~} \frac{ Sp(n) } {SU(n) \times U(1)} ~~,
{~~~~~~~~~~~~~~~~~~}
$$
$$
{~~~~~~~~} \frac{ SO(2n) }{SU(n) \times U(1) } ~~, {~~~~~~~}
\frac{ SO(m + 2) }{ SO(m) \times SO(2)} ~~, \quad m>2
 \eqno(12) $$
for which the equations (11) can be solved exactly, 
according to the rules given in \cite{gk}.  The specific feature 
of the compact K\"ahler symmetric spaces is that eqs. (11)
are equivalent to the holomorphic geodesic equation
$$ \frac{ {\rm d}^2 \U^I_\star (w) }{ {\rm d} w^2 } \,\,+ \,\,
\G^I_{JK} \left( \U_\star (w), \bar{\F} \right)
\frac{ {\rm d} \U^J_\star (w) }{ {\rm d} w }
\frac{ {\rm d} \U^K_\star (w) }{ {\rm d} w }  ~=~ 0 
\label{geo1}
\eqno(13) $$
under the initial conditions (10). Here $\G^I_{JK} 
( \F , \bar{\F} )$ are the Christoffel symbols for the  
K\"ahler metric $g_{I \bar J} ( \F , \bar{\F} ) = \pa_I 
\pa_ {\bar J}K ( \F , \bar{\F} )$.

Upon elimination of the auxiliary superfields, the action 
(4) takes the form
$$  
S  [\U_{\star}, \breve{\U}_{\star}] ~=~ S_{{\rm tb}}[\F, \bar 
\F, \S, \bar \S]  ~=~ \int {\rm d}^8 z \, \Big\{\,
K \big( \F, \bar{\F} \big) ~-~ g_{I \bar{J}} \big( \F, \bar{\F} 
\big) \S^I {\bar \S}^{\bar{J}} \, {~~~~~~~~~~~~~~~~~~~~~~~~~
~~~~~~~~~~~~~~~~~~~~}  $$
$${~~~~~~~~~~~~~~~~~~~~~~~~~~~~~~~~~~~~~~~~~~~~~}+~
\sum_{p=2}^{\infty} \cR_{I_1 \cdots I_p {\bar J}_1 \cdots {\bar 
J}_p }  \big( \F, \bar{\F} \big) \S^{I_1} \dots \S^{I_p} {\bar 
\S}^{ {\bar J}_1 } \dots {\bar \S}^{ {\bar J}_p }~\Big\}~~~, 
\eqno(14)  $$
where the tensors $\cR_{I_1 \cdots I_p {\bar J}_1 \cdots {\bar 
J}_p }$ are functions of the Riemann curvature $R_{I {\bar 
J} K {\bar L}} \big( \F, \bar{\F} \big) $ and its covariant 
derivatives.  Each term in the action contains equal powers
of $\S$ and $\bar \S$, since the original model (4) 
is invariant under rigid U(1)  transformations
$$ 
\U(w) ~~ \longrightarrow ~~ \U({\rm e}^{{\rm i} \a} w) 
\quad \Longleftrightarrow \quad 
\U_n(z) ~~ \longrightarrow ~~ {\rm e}^{{\rm i} n \a} \U_n(z) 
~~~.
\label{rfiber}
\eqno(15) $$

To get a better feel for this construction, let us consider the 
simple example of ${\Bbb C}P^n = SU(n+1) / U(n)$ in the role 
of the K\"ahler manifold $\cM$. For ${\Bbb C}P^n$ we have 
$$ K (\F, {\bar \F}) = r^2 \ln \left(1 + \frac{1}{r^2}  
\F^L \overline{\F^L} \right)
~,~~~
g_{I {\bar J}} (\F, \bar \F) =  
\frac{ r^2 \d_{I {\bar J}} }{r^2 + \F^L \overline{\F^L} }
- \frac{  r^2   \overline{\F^I} \F^J  }
{(r^2 + \F^L \overline{\F^L})^2 }
\label{s2pot}
\eqno(16) $$
with $1/r^2$ being proportional to the curvature of  ${\Bbb C}P^n$.
Direct calculations lead to  
$$ S[\U_\star, \breve{\U}_\star] ~=~ \int {\rm d}^8 z \, \left\{ ~
K(\F, \bar \F) ~+ ~ r^2 \ln \Big(1 - \frac{1}{r^2} \, g_{I {\bar J }} 
(\F, {\bar \F})\; \S^I {\bar \S}^{\bar J} \Big) \right\} ~.
\label{physpol}
\eqno(17) $$
As is seen, the action is well defined when $|\S |^2 \equiv g_{I 
{\bar J}} \; \S^I {\bar \S}^{\bar J} < r^2$.  Under this restriction 
we can represent the second term in (17) by a Taylor series in $| 
\S |^2$, and the series  is nothing but an expansion in powers of
the curvature of ${\Bbb C}P^n$.

We can turn this to our advantage to obtain some partial information 
about the tensors $\cR_{I_1 \cdots I_p {\bar J}_1 \cdots {\bar 
J}_p }$ that appear in (14). Since the curvature of ${\Bbb C}P^n$ is
covariantly constant, the expansion of (17) fixes the form of all the 
terms in $\cR_{I_1 \cdots I_p  {\bar J}_1 \cdots {\bar  J}_p }$ that 
are not dependent of any derivatives of the K\" ahler manifold curvature.
Appealing to universality, we suggest that these tensors in (14) 
should not be strongly dependent on the choice of a particular K\" ahler
manifold. By this assertion, it follows that all the non-derivative
terms in $\cR_{I_1 \cdots I_p {\bar J}_1 \cdots {\bar J}_p }$ are
fixed by the series expansion of the logarithm in (17). Of course,
a really comprehensive understanding of this opproach requires a 
completely explicit description of $\cR_{I_1 \cdots I_p {\bar J}_1 
\cdots {\bar J}_p }$ which presently lies beyond our grasp. This
a topic for future investigation.

The Lagrangian of the $N=2$ supersymmetric model (14) cannot yet be 
identified with a hyperk\"ahler potential. The point is that for $N=1$
rigid supersymmetric models  their Lagrangians coincide with K\"ahler
potentials only if  all the dynamical variables are chiral superfields.  
But our model (14) is described by chiral superfields $\F^I$ 
(parametrizing the base K\"ahler manifold) and by complex linear 
ones $\S^I$ (parametrizing the tangent fibers). It remains, however, 
to fulfil one more step -- to dualize the linear superfields $\S^I$ 
into chiral ones $\J_I$, ${\bar D}_\ad \J_I = 0$, via the Legendre 
transform
$$ S_{{\rm tb}}[\F, \bar \F, \S, \bar \S] ~ \rightarrow ~
S[\F, \bar \F, U, \bar U, \J, \bar \J] ~=~ S_{{\rm tb}}[\F, 
\bar \F, U, \bar U] ~-~  \int {\rm d}^8 z \,  \Big\{   
 U^I \J_I ~+~ {\rm c.c.} \Big\}
\label{auxmodel}
\eqno(18) $$
with the auxiliary superfields $U^I$ being complex unconstrained. 
By construction, $\{U^I\} $ is a  tangent vector at point $\F$ of
$\cM$, therefore $\{ \J_I \}$ is a one-form at the same point.
Eliminating the auxiliary variables $U^I$ and ${\bar U}^{\bar J}$
in (18), with the aid of their equations of motion,
results in a purely chiral sigma model $S_{{\rm cb}}[\F, \bar \F, 
\J, \bar \J] $ which is dually equivalent to the $N=2$ supersymmetric 
model (14) and is defined on the cotangent bundle $T^{\star}
\cM$.   Therefore, the superfield  Lagrangian for $S_{{\rm cb}}[\F, 
\bar \F, \J, \bar \J] $ coincides  with the hyperk\"ahler potential 
of $T^{\star} \cM$.  In particular,  if one applies the Legendre 
transform described to the model (\ref{physpol}), one exactly reproduces 
the hyperk\"ahler potential on the complete cotangent bundle $T^{\star}
({\Bbb C}P^n)$ \cite{calabi, perelomov}, see \cite{gk} for more 
details.

In conclusion, we would like to point out that the $c$-map 
hyperk\"ahler potential (2) is generated by a self-coupling 
of $N=2$ tensor multiplets.  As concerns the hyperk\"ahler structures 
on the cotangent bundles of K\"ahler manifolds, the above consideration
shows these are generated by self-couplings of $N=2$ polar multiplets.

\vskip0.5cm
\noindent
{\large \bf Acknowledgements}

\smallskip
\noindent
We like to thank I. Buchbinder, B. de Wit, N. Dragon, F. Gonzalez-Rey, 
T. H\"ubsch, E. Ivanov,  U. Lindstr\"om, A. Lossev
D.  L\"ust, A. Nersessian,  B. Ovrut, M. Ro\v{c}ek, S.
Theisen and M. Vasiliev for enlightening discussions.
SMK is grateful to the organizers of Buckow-98
for financial support. The work of SJG was
supported in part by NSF Grant PHY-98-02551.
The work of SMK was supported in part by 
Deutsche Forschungsgemeinschaft.

\end{document}